\documentclass[sigconf]{acmart}
\usepackage[utf8]{inputenc}
\usepackage{booktabs} % For formal tables
\usepackage{xspace}
\usepackage{tikz}
\usepackage{balance}
\usetikzlibrary{calc}
\newcommand{\xlap}{\textsc{X-Lap}\xspace}

\setcopyright{rightsretained}

\acmDOI{https://doi.org/10.1145/3314206.3314211}

\acmISBN{}

\acmConference[ECRTS-RTN'18]{RTN'2018 The 16th International Workshop on Real-Time Networks}{July 2018}{Barcelona, Spain}
\acmYear{2018}
\copyrightyear{2018}

\acmPrice{}

\begin{document}
\title{$\Delta$\textsc{elta}: Differential Energy-Efficiency, Latency, and Timing Analysis for Real-Time Networks}

\author{Stefan Reif}
\affiliation{\institution{Friedrich-Alexander University Erlangen-N\"urnberg}}
\email{reif@cs.fau.de}

\author{Andreas Schmidt}
\affiliation{\institution{Saarland Informatics Campus}}
\email{andreas.schmidt@cs.uni-saarland.de}

\author{Timo Hönig}
\affiliation{\institution{Friedrich-Alexander University Erlangen-N\"urnberg}}

\email{thoenig@cs.fau.de}

\author{Thorsten Herfet}
\affiliation{\institution{Saarland Informatics Campus}}
\email{herfet@cs.uni-saarland.de}

\author{Wolfgang Schröder-Preikschat}
\affiliation{\institution{Friedrich-Alexander University Erlangen-N\"urnberg}}
\email{wosch@cs.fau.de}

\renewcommand{\shortauthors}{S. Reif et al.}

\begin{abstract}
    The continuously increasing degree of automation in many areas~(e.g. manufacturing engineering, public infrastructure) lead to the construction of cyber-physical systems and cyber-physical networks. To both, time and energy are the most critical operating resources. Considering for instance the Tactile Internet specification, end-to-end latencies in these systems must be below 1ms, which means that both communication and system latencies are in the same order of magnitude and must be predictably low. As control loops are commonly handled over different variants of network infrastructure~(e.g. mobile and fibre links) particular attention must be payed to the design of reliable, yet fast and energy-efficient data-transmission channels that are robust towards unexpected transmission failures. As design goals are often conflicting~(e.g. high performance vs. low energy), it is necessary to analyze and investigate trade-offs with regards to design decisions during the construction of cyber-physical networks.

In this paper, we present $\Delta$elta, an approach towards a tool-supported construction process for cyber-physical networks. $\Delta$elta extends the previously presented \xlap tool by new analysis features, but keeps the original measurements facilities unchanged. $\Delta$elta jointly analyzes and correlates the runtime behavior~(i.e. performance, latency) and energy demand of individual system components. It provides an automated analysis with precise thread-local time interpolation, control-flow extraction, and examination of latency criticality. We further demonstrate the applicability of $\Delta$elta with an evaluation of a prototypical implementation.

\end{abstract}




\hyphenation{time-stamp time-stamps cyc-le-stamp cyc-le-stamps}

\maketitle

\hypersetup{
pdftitle={Delta: Differential Energy-Efficiency, Latency, and Timing Analysis for Real-Time Networks}
}

\section{Introduction}
\label{sec:introduction}

With the increasing degree of automation in domains such as manufacturing engineering, mobility and logistics, as well as public infrastructure, great efforts are being made to close the gap between cyber (or digital) and physical processes, commonly described as \textit{cyber-physical systems}~(CPS).
The communication between these CPSs is becoming particularly important, as control loops are handled over different variants of network infrastructure (e.g. fibre links, LTE/5G) to make measurements (i.e. sensor data) available from a variety of systems and implement the necessary, distributed coordination. These communicating systems are commonly referred to as \textit{cyber-physical networks}~(CPN) and represent a special division of   \textit{real-time networks}. CPNs share the requirements of the underlying cyber-physical systems which implies that such networks must ensure fault-tolerance or resilience, provide real-time characteristics, allow the execution of distributed tasks, and incorporate self-sufficient systems which are battery driven and demand low-power operations with predictable energy footprints.

CPNs not only require novel solutions on all layers of a communication stack (i.e. transport, network, operating system and hardware), they also demand for cross-layer integrations to fulfill these requirements~\cite{lee2008cyber}. The work on solutions for the arising challenges requires tools that analyze the timing behavior as well as the energy demand at system level. Correspondingly, such development tools support the construction process of CPNs and allow the evaluation of system changes, validate guarantees~(i.e. as to time and energy demand), and guide the process of decision making for further improvements of the system design. This paper presents and evaluates an approach towards tool-based, automated timing and energy demand analysis of CPNs. The approach improves and extends \xlap~\cite{larn:2017:rtn}, a timing-analysis tool specifically designed for CPNs.
We demonstrate the validity of the proposed approach by analyzing the \textit{Predictably Reliable Real-time Transport~(PRRT)} protocol, which is a transport layer protocol for CPNs. PRRT respects application constraints, such as maximum latency and tolerance to lost messages, and parametrizes its internal mechanisms for congestion and error control to fulfill these.

The contribution of this paper is threefold:

\begin{itemize}
    \item We present an approach which jointly captures, analyzes, and correlates the runtime behavior~(i.e. performance, latency) and energy demand of CPNs.
    \item We propose an automated analysis which provides precise thread-local time interpolation, control-flow extraction, and analysis of latency criticality.
    \item We extend our analysis to compare multiple sets of traces, which automatically extracts the impact of code changes or hardware configurations (i.e. varying processor speeds).
\end{itemize}

The remainder of this paper is structured as follows: Section~\ref{sec:background} gives background information as well as related work. Section~\ref{sec:system} presents details on the implementation, while Section~\ref{sec:evaluation} shows the evaluation results. Section~\ref{sec:conclusion} concludes the paper and gives directions for further research and implementation.

\section{Background and Related Work}
\label{sec:background}

In CPSs and CPNs in particular, we see that the requirements that are posed~\cite{lee2008cyber} cannot be fulfilled by analyzing the network or operating system independently. However, the networking and operating systems domains have been using completely different sets of tools to evaluate system performance and validate adherence to design goals. Typically, this process involves abstracting away the other domain~\cite{Schimmel:IEC,Ferrandiz:2009} and simplifying it by characterising representative loads (from a system perspective) or representative task execution times (from a network perspective). With the precision and timescale required for CPSs~\cite{barroso:2017:cacm}, these abstractions are no longer sufficient and it is required to properly co-design and analyze both domains at the same time~\cite{beckman:2006:iccc,dean:2013:cacm}.

Communication and operating system latencies used to be in different orders of magnitude (communication: 10ms to 10s; system: 10us to 1ms), if we for instance consider applications running on the Internet and covering large physical distances. In contrast, CPN applications have stricter time constraints (cf. the Tactile Internet~\cite{fettweis2014tactile}), e.g. a total end-to-end latency of 1ms, so that the two domains tend to contribute to the overall latency more evenly. For the networking domain, this means that the communication distances are strictly limited (below 300km to achieve 1ms propagation delay with speed of light) and queueing delays must be kept as close to zero as possible, using pacing and congestion control. The latter is already motivated in previous work on data center networking~\cite{alizadeh2012less}, which does not deal with real-time requirements, but still requires that the delays are predictable and low~\cite{dean:2013:cacm}. Considering that data centers and clouds are going to play an import role in offloading tasks for upcoming CPN applications, these approaches will also be part of network stacks tailored to the needs of CPNs.

Another approach to timing and energy-demand analysis is off-line static analysis, before run-time. The goal of such an analysis is typically to derive a \emph{worst-case execution time} (WCET)~\cite{wilhelm:2008:acmtecs} and \emph{worst-case travel time} (WCTT)~\cite{Liu:2016} for network protocols. $\Delta$elta and \xlap, in contrast, monitor the behavior at run-time. In consequence, $\Delta$elta is inherently restricted to the actually observed behavior, and it cannot guarantee that the collected samples cover worst-case scenarios. However, this approach is still viable for soft real-time systems that are often too complex for exhaustive static analysis. Besides, the information gathered with \xlap and $\Delta$elta can be used to verify timing information gained by static analysis.
An example for a tool-based approach for the construction of real-time networks based on static analysis is RT-Appia~\cite{rodrigues2001design}. This framework composes protocol components to real-time network stacks. Thus, the protocol stack is tailored to the application demand in order to improve WCET analyzability.

\subsection{\xlap}
\xlap\footnote{\url{http://xlap.larn.systems}}~\cite{larn:2017:rtn} is a timing analysis tool specifically designed for CPNs. In these networks, the network protocol stacks are often tightly coupled with operating system in order to fulfill real-time requirements with high reliability. \xlap therefore works cross-layer, by profiling the performance on transport, operating system and network layer. Furthermore, it works on multiple nodes and allows to correlate the timing information taken on these systems, without explicit clock synchronization. \xlap consists of two components: 1) a set of C-level calls to be injected into the code for capturing high resolution clock- and cycle-stamps with minimal overhead; 2) a set of analysis procedures to determine the causes of latency and jitter. Sender and receiver systems that use these C-level calls are going to produce a table with packet traces after termination, which gives the clock- and cycle stamps together with the packet sequence number. Afterwards, the processing chain of \xlap interpolates timestamps---we only take them when necessary, because cycle-stamps lead to a lower overhead induced by the profiling---and calculates the durations of certain processing steps, as well as the overall execution time by different threads. Consequently, \xlap is able to compute exact processing durations for fine-grained processing steps of every packet, through the entire CPN---providing insights that other (single-layer) tools, e.g. \texttt{wireshark}\footnote{https://www.wireshark.org/} for networking or \texttt{valgrind}\footnote{http://valgrind.org/} for systems, cannot deliver.

\subsection{Predictably Reliable Real-time Transport}
We use \xlap for evaluating the \textit{Predictably Reliable Real-time Transport~(PRRT)}\footnote{\url{http://prrt.larn.systems}} protocol~\cite{Gorius:PRRT}. PRRT is a protocol that aims to overcome the shortcomings of existing transport layer protocols, such as TCP, UDP and QUIC~\cite{QUIC}, in particular with respect to time- and resilience-awareness. In contrast to these solutions, PRRT provides a \textit{partially reliable ordered datagram stream with bounded latency}, in contrast to TCP~(\textit{fully reliable ordered byte stream}), UDP~(\textit{datagram stream}) and QUIC~(\textit{multiplexed fully reliable ordered byte streams with reduced number of round-trips}), which all do not provide bounds on timing. PRRT achieves this by a) letting the application state its requirements on maximum latency and acceptable residual error rate and b) incorporating this into the protocol's operation, together with measurements of the channel's propagation delay, loss rate and bottleneck bandwidth. Therefore, it is necessary that code segments across multiple threads within PRRT take a predictable amount of time and face only minimal jitter. For achieving predictable delay on the network layer, PRRT uses multiple mechanisms: First, it uses \textit{adaptive hybrid error correction}~\cite{Lee:Throughput} to chose and optimal configuration for providing proactive~(FEC) and reactive~(ARQ) error control, based on the application constraints and channel parameters. An optimal coding scheme ensures that after the application's maximum tolerable latency, at most it's maximum tolerable residual loss is present. Second, it uses packet pacing to adjust its effective sending rate to the bottleneck bandwidth, effectively avoiding queueing delays when packets are send out in bursts. Finally, pacing together with measurements on the propagation delay are used to compute the congestion window over the given path and avoid further losses due to competitions for bandwidth with other applications.

In general, evaluations using \xlap can be executed with other transport layer protocols as well, as long as the source code is available and can be changed to inject clock- as well as cyclestamping calls. Furthermore, the protocol must provide some notion to identify packets, e.g. a sequence number, so that traces for individual pieces of data can be generated.

\section{Implementation}
\label{sec:system}

The implementation of $\Delta$elta bases on \xlap~\cite{larn:2017:rtn}, and reuses its high-precision, low-interference time-stamping architecture. However, it significantly extends the off-line analysis component, providing additional insights and increasing automation.

The analysis approaches we present in this paper augment the timing analysis facilities of \xlap by a \emph{differential} approach that \emph{compares} the timing of multiple experiments. We extract information that are particularly useful for the design, implementation, and evaluation of real-time networks.

The analysis approaches work \emph{retroactively}, in the sense that they derive information from observing a real-time network stack in action. Such information is necessarily imprecise because it is restricted to the actually observed behavior. However, we show in the following, that we can derive valuable information with little design-time and run-time costs. \emph{Proactive} approaches that \emph{predict} the behavior, for instance by static analysis, are beyond the scope of this paper.

\subsection{Control Flow Reconstruction}
\label{sec:control_flow_analysis}
\begin{figure*}
    \includegraphics[width=1\textwidth]{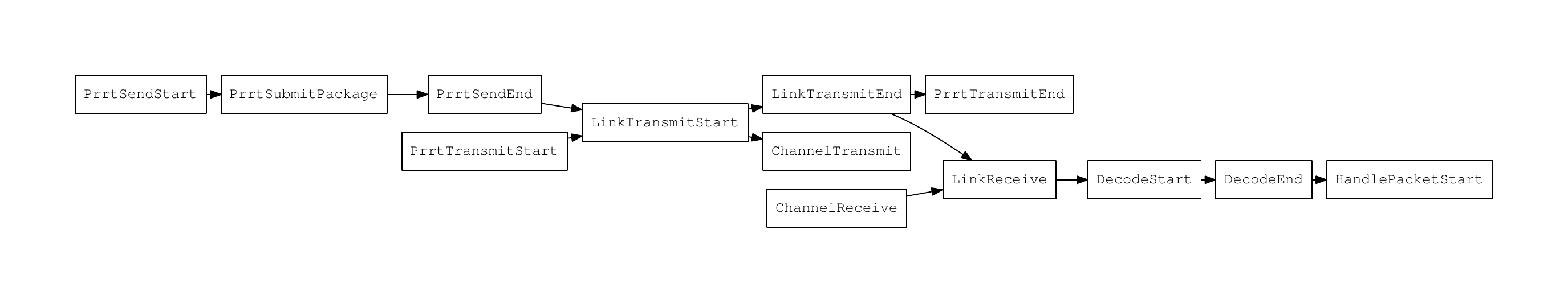}
    \vspace{-1.2cm}
    \caption{A control flow graph reconstructed from the \emph{happens-directly-before} relation}
    \label{fig:graph}
\end{figure*}

When searching for and mitigating the interferences between different processing steps in a transport stack, the control flow graph is an essential tool to foster the analysis. As control flows can vary between different protocol versions or even minimal code changes, it is crucial for a cross-layer analysis to extract this piece of information from the captured data. Obviously, this cannot fully replace the expertise of a developer and a thorough analysis of the source code, but in particular with concurrent systems, such an empirical approach can greatly support the development process.

The reconstruction works as follows: We compute a \emph{happens-before} relation~\cite{lamport:1978:cacm} $A \rightarrow^{+} B$ of events $A$ and $B$ if, for every packet, $A$ has a lower time-stamp than $B$. If $A \not\rightarrow^{+} B$ and $B \not\rightarrow^{+} A$, $A$ and $B$ are concurrent. Concurrency can occur if events happen in different threads, or in interrupt handlers, such as timers.

Since \emph{happens-before} is transitive, we derive a \emph{happens-directly-before} relation $A \rightarrow B$ of events $A$ and $B$ if $A \rightarrow^{+} B$ and no event $E$ exists with $A \rightarrow^{+} E \rightarrow^{+} B$. This relation reconstructs control flows, because it reveals the sequence of events in the protocol stack. A representative sub-graph is depicted in Figure~\ref{fig:graph}, showing the flow packets follow through a specific version of the PRRT stack.

\subsection{Latency-Criticality Analysis}
The analysis steps in the following sections rely on the information whether the duration of a  specific code segment impacts the end-to-end latency. We therefore need to quantify the relation between code segments and the end-to-end latency. Thereto, we define the \emph{latency criticality} of a code segment $\left<A,B\right>$, as the Pearson correlation coefficient between the duration of $\left<A,B\right>$ and the \emph{end-to-end} (E2E) latency, over all packets in a trace. This value describes whether the E2E latency depends on the duration of a code segment. Thereby, we utilize the control flow reconstruction to automatically detect whether $A$ and $B$ refer to an actual code segment in the protocol stack.

The latency-criticality analysis measures how individual code segments influence the end-to-end delay. Thus, it gives a hint on which parts of the protocol are worth considering for further optimization, to fulfill the requirements of CPNs.

\subsection{Timing-Predictability Analysis}
\label{sec:timing_predictability_analysis}

A key property of CPNs, as well as real-time networks, is timing predictability. Therefore, our analysis identifies protocol parts that exhibit unpredictable behavior.

The reasons for non-reproducible timing behavior are many-fold. For instance, many network protocols incorporate concurrency due to timers, task synchronization or thread interaction. Besides, hardware and software-related interference such as cache misses or OS noise~\cite{tsafrir:2005:ics} influence the timing behavior of code segments. Furthermore, the timing of specific code segments might depend on current channel properties (e.g. packet loss or propagation delay).

In order to identify code parts where the timing behavior is not sufficiently predictable, we run the same exact version of the code multiple times and trace each packet using \xlap. Afterwards, we reconstruct the control flow from the timestamps to identify the code segments of interest.

For each code segment, we apply the \emph{k-sample Anderson-Darling-Test}~\cite{scholz:1987:adtest}. This stochastic test is a well-established method to check whether multiple samples are derived from the same distribution. While, technically, the durations are drawn from the same actual runtime-distribution of code segments, the actual data might show significant differences between consecutive runs.

The stochastic test only classifies timing behavior as ``different'' when it is certain, given a configurable significance parameter. When the test is not certain enough, it classifies the timing distributions as ``similar''. However, this test can give false-negatives when timing is actually unpredictable but, incidentally, behaves the same at all observed experiment runs. However, if the test result is ``different'', then the two experiment runs have certainly shown different timing behavior.

The actual importance of the unpredictability results depends on the latency criticality of the code segment. At some protocol parts, unpredictability can be expected, for instance, regarding timers that cause concurrent code execution. Therefore, the timing unpredictability results should be analyzed jointly with the results of the latency-criticality analysis.

\subsection{Modification Tracking}

When changing parts of the implementation of protocols, the timing behavior of (at the first glance) independent protocol parts can change. The reason is interference that is sometimes hidden: For instance, cache effects and contention on hardware buses can cause non-trivial interference between seemingly independent protocol parts. Besides, if the timing of network packets changes slightly, the entire protocol behavior might adapt.

Our analysis reveals such timing interference. We execute different protocol versions and, for each version, capture packet traces with \xlap. Then, we apply the Anderson-Darling-Test similar to the timing predictability evaluation described in Section~\ref{sec:timing_predictability_analysis}. This test reveals code segments where the timing behavior has changed. We can also identify control flow changes using the automated control flow reconstruction described in Section~\ref{sec:control_flow_analysis}.

\subsection{Energy-Efficiency Analysis}

Another use of differential analysis for real-time networks is energy-efficiency optimization. Modern hardware components offer multiple ``knobs'' that affect power demand as well as performance. These configurations include processor frequencies and sleep states with varying wake-up latencies. Identification of the most efficient states while maintaining desired real-time behavior is therefore a complex problem~\cite{miyoshi:2002:ics}.

To analyze the relation between energy demand and performance, we trace packets with \xlap and measure the end-to-end latency. Simultaneously, we measure the energy demand of the sender and the receiver hosts during the experiment. We repeat these measurements for the various configurations of the hardware and thereby identify the most energy-efficient setup.

\subsection{Slowdown Analysis}

Besides finding the most energy-efficient hardware configuration, $\Delta$elta offers additional fine-grained timing information, giving insight into the detailed behavior of individual protocol parts.

To analyse how protocol parts behave under different hardware power states, we measure a \emph{fast} and a \emph{slow} experiment run. Then, we reconstruct the control flow and, for each code segment $X$, we compute the average delays $D_{X,fast}$ and $D_{X,slow}$, from the fine-grained timing information of the respective experiment runs. We compute the \emph{slowdown} $S_{X} $of each segment, and we normalize the slowdown values using the end-to-end slowdown:

\begin{equation*}
  S_{X} = \frac{D_{X,slow}}{D_{X,fast}} \qquad S_{X}^{*} = \frac{S_{X}}{S_{E2E}}
\end{equation*}

The normalized slowdown reveals whether the duration a code segment depends on the processor speed. If $S_{X}^{*}$ is above $1$, the segment latency suffers from the slow hardware configuration. A value below $1$, however, means that the performance loss at a code segment is less than the end-to-end performance loss.

\section{Evaluation}
\label{sec:evaluation}

We evaluate our analysis approaches using \xlap and PRRT. We obtain detailed package traces and submit them to the $\Delta$elta analysis techniques described in Section~\ref{sec:system}. The purpose of this evaluation is to find out whether the analysis techniques presented in Section~\ref{sec:system} produce meaningful results. We therefore apply them to a real-time networking protocol, PRRT.

\subsection{Experimental Setup}

We run all experiments on two nodes in a dedicated networking testbed to make sure that only minimal interference on the hosts and the network disturb our experiments.

\subsection{Control Flow Reconstruction}

To evaluate the control flow reconstruction, we have traced 4095 packets with \xlap. Figure\,\ref{fig:graph} depicts a representative sub-graph of the computed \emph{happens-directly-before} relation. For events that happen in the same thread, the analysis reconstructs the control flow reliably. However, one computed flow was a \emph{false-positive} because, by chance, an event in one thread always happened before another event in another thread. We also encountered one \emph{false-negative}, where two events happened nearly simultaneously and the clock resolution was not sufficient to order the two timestamps.

\subsection{Latency Criticality Evaluation}

\begin{figure}
 \includegraphics[scale=0.5]{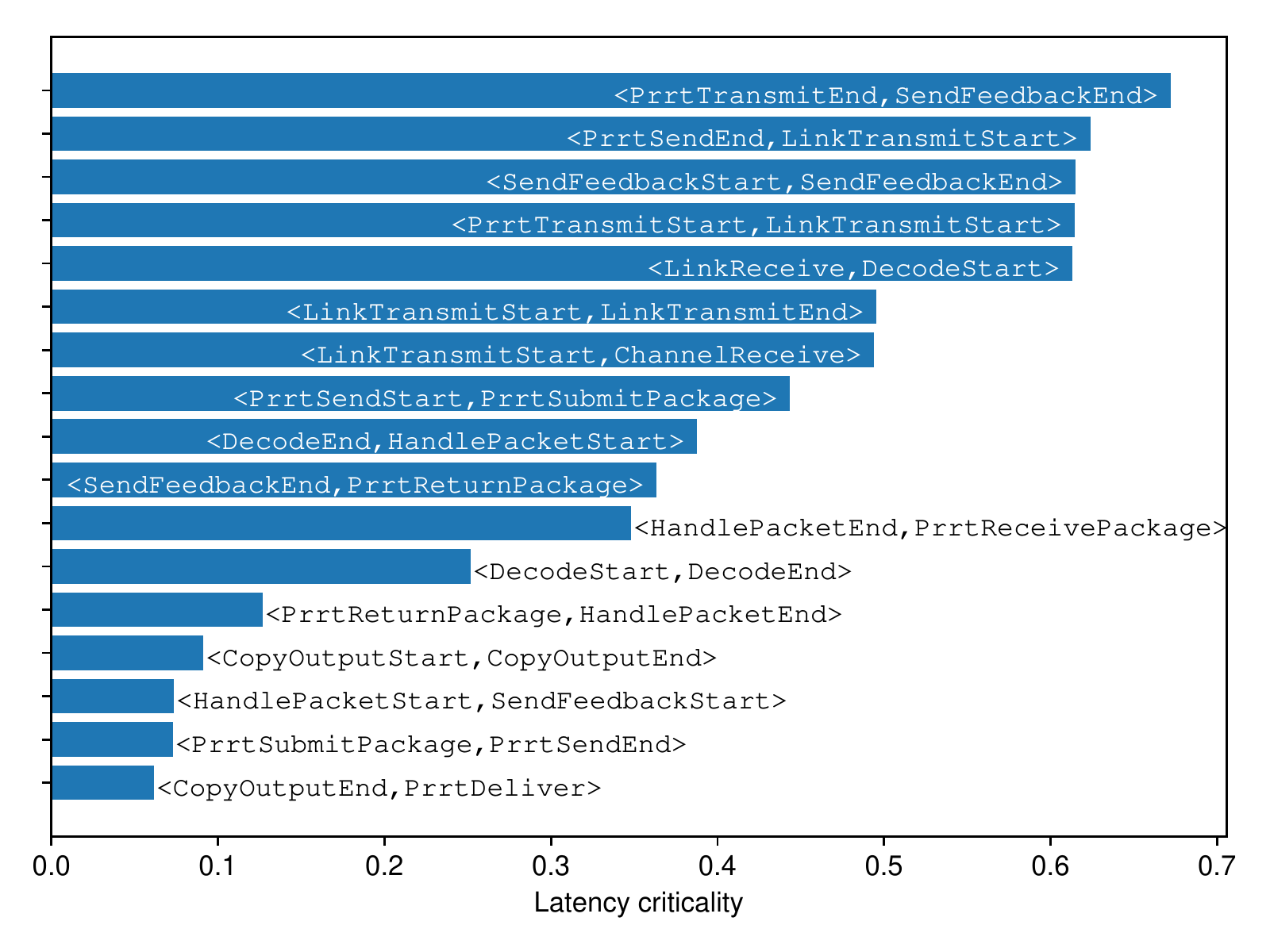}
 \caption{Latency criticality of protocol parts}
 \label{fig:eval:latency_criticality}
\end{figure}

We evaluate the latency criticality of all code segments identified by the control flow reconstruction. Figure~\ref{fig:eval:latency_criticality} visualizes results for the most-critical code segments. The values correspond loosely to the average segment latency, because jitter is often proportional to the duration. Therefore, long segments correlate stronger with the end-to-end latency. However, some segments have low criticality because of concurrency in the protocol stack.

\subsection{Timing Predictability Evaluation}

To evaluate the timing predictability analysis, we have traced two samples with 100 packets each. Both samples use identical code revisions, and the hardware frequency was fixed to 3\,GHz. We run the Anderson-Darling-Test on both data sets, which reveals that $9$ code segments have non-reproducible timing, $5$ of which were expected because of concurrency or interaction with the network hardware. Out of the $4$ remaining segments, only one has a relatively high latency criticality (\texttt{<PrrtSendStart,PrrtSubmitPackage>}). Thus, the analysis gives a hint where the protocol implementation can be optimized to improve timing predictability.

\subsection{Modification Tracking and Evaluation}

We repeat the timing predictability analysis, but we add an artificial \emph{sleep} call as a dummy code modification that impacts the timing behavior of a single code segment. The sleep call was already in a segment with non-reproducible timing, at the sender side. We compare the results with the timing reproducibility analysis to eliminate segments where we know that the timing changes between experiment runs. The analysis reveals that two code segments at the receiver side change their timing behavior, according to the Anderson-Darling-Test. Thus, our analysis reveals interference between protocol parts.

\subsection{Energy Efficiency}

\begin{figure}
 \includegraphics[scale=0.5]{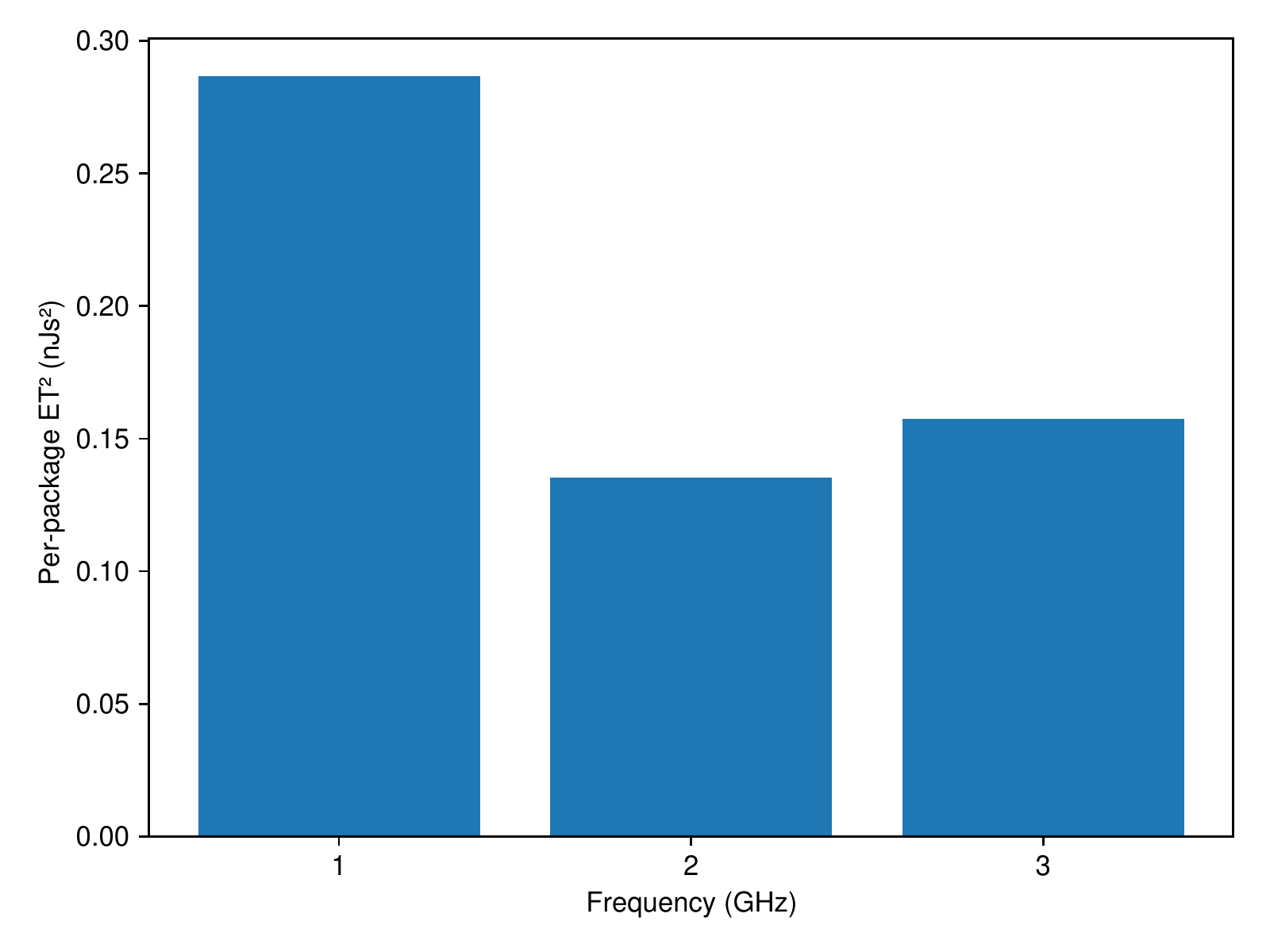}
 \caption{Energy efficiency at different processor speed settings (lower is better)}
 \label{fig:eval:energy_efficiency}
\end{figure}

We evaluate the system energy efficiency at three performance states: We configure our system to run at a fixed speed of 1\,GHz, 2\,GHz, or 3\,GHz, for each evaluation run. Thereby, we use \xlap to measure the end-to-end package latency. Simultaneously, we measure the energy demand with RAPL~\cite{intel:ia:sdm}.

We compute the average ET$^2$ metric~\cite{martin:2002:et2} per packet, because it is a fair comparison for DVFS settings. Figure~\ref{fig:eval:energy_efficiency} summarizes the evaluation results. Thereby, the most energy-efficient configuration is neither the fastest nor the slowest. This result is well-aligned with other research~\cite{lesueur:2011:atc}. This outcome also indicates that energy efficiency needs actual energy measurements to find the optimal hardware configuration.

\subsection{Normalized Slowdown}

\begin{figure}
 \includegraphics[scale=0.5]{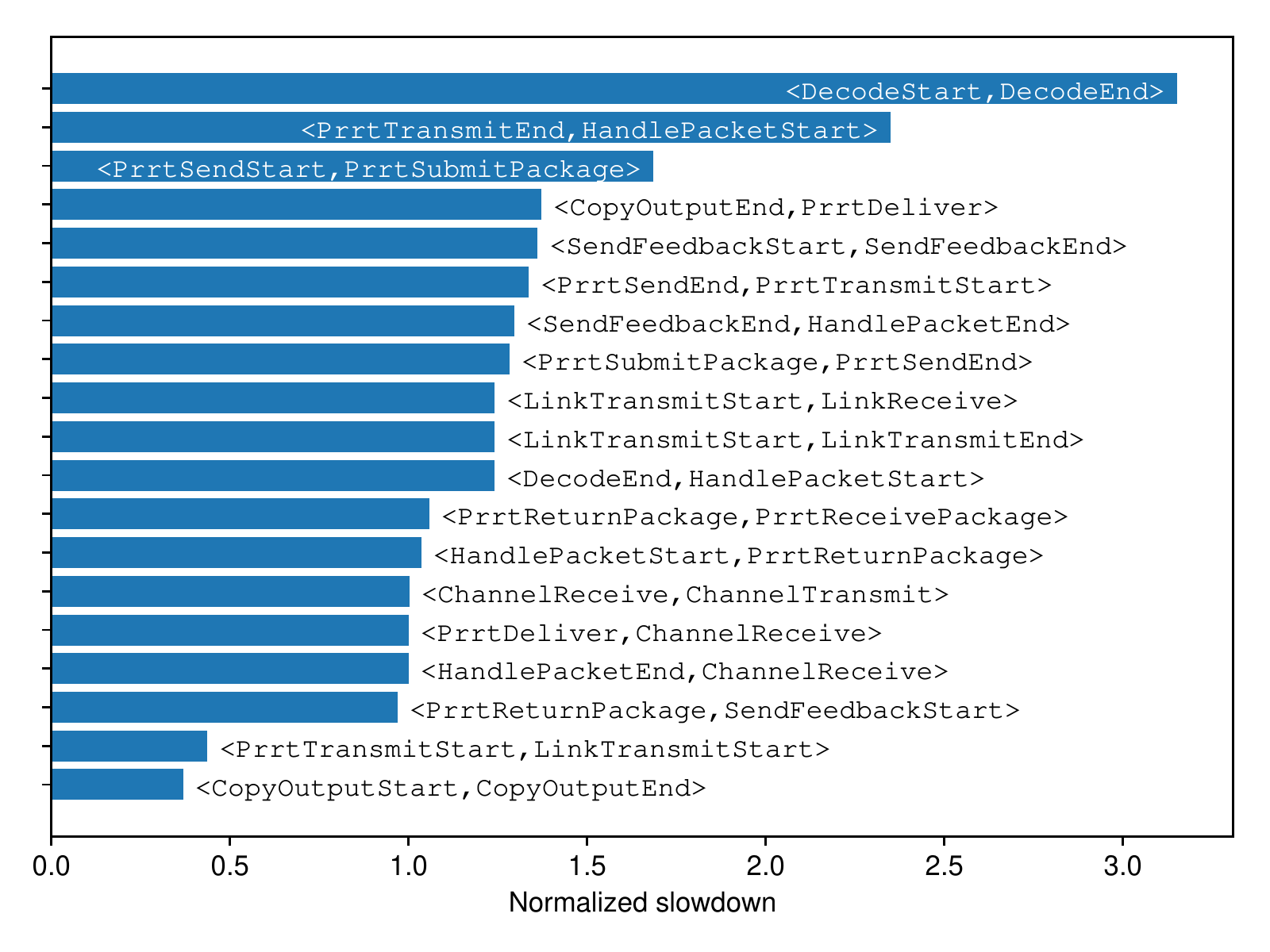}
 \caption{Normalized slowdown for code segments}
 \label{fig:eval:slowdown}
\end{figure}

We further analyse the precise packet timings at 2\,GHz and 3\,GHz and compute the normalized slowdown for each code segment, using the control flow reconstruction to identify relevant code segments. The results are summarized in Figure~\ref{fig:eval:slowdown}, showing that on the one hand, we find a relatively high normalized slowdown at the \texttt{<DecodingStart,DecodingEnd>} segment, because the \emph{decoding} operation is CPU bound. On the other hand, memory-bound segments, such as \texttt{<CopyOutputStart,CopyOutputEnd>}, exhibit a low normalized slowdown. These results indicate that different error codes (or optimized versions of it) might allow to use a more energy-efficient hardware configuration.

\section{Conclusion}
\label{sec:conclusion}

This paper has introduced $\Delta$elta,  a collection of analysis techniques for cyber-physical networks. In summary, these novel techniques offer detailed insight on the fine-grained timing behavior and energy demand.
We automatically reconstruct control flows from time measurements. We utilise this information in a differential analysis approach to detect code segments with unpredictable timing. Additionally, our analysis detects the impact on code revisions on the overall system behavior. The evaluation shows that minor changes in one processing step can indeed change the timing of various other, seemingly unrelated, parts of the protocol.
We further combine precise timing information with actual energy measurements. Our analysis identifies the most energy-efficient hardware configuration, and evaluates quantitatively which code segments tolerate low processor speeds.

In summary, our proposed approach helps designers of cyber-physical networks to verify timing properties, to reduce the end-to-end latency and jitter, and to increase the energy efficiency of the overall systems.

In future work, we are going to apply these analyses to improve transport protocol stacks for cyber-physical networks, to achieve predictably low end-to-end latency and a high energy efficiency.

\begin{acks}
    The work is supported by the \grantsponsor{DFG}{German Research Foundation~(DFG)}{TODO} as part of SPP 1914 ``Cyber-Physical Networking'' under grants \grantnum{DFG}{HE~2584/4-1} and \grantnum{DFG}{SCHR~603/15-1}.
\end{acks}

\balance
\bibliographystyle{ACM-Reference-Format}
\bibliography{ms}

\end{document}